\documentstyle[12pt,a41,epsfig]{article}
\begin{document}
\thispagestyle{empty}
\begin{flushright}
\large
CERN-TH/97-287 \\
DTP/97/94 \\
October 1997
\end{flushright}
\vspace{1.2cm}

\begin{center}
\LARGE
{\bf Polarized Lambda Production at HERA} 

\vspace*{1.8cm}
{\Large D.\ de Florian$^a$, M.\ Stratmann$^b$, W.\ Vogelsang$^a$ }

\vspace*{1.5cm}
\large
{\it $^a$Theoretical Physics Division, CERN, CH-1211 Geneva 23, Switzerland}

\vspace*{0.1cm}
{\it $^b$Department of Physics, University of Durham, Durham DH1 3LE, England}

\vspace{2.6cm}
{\bf Abstract} 
\end{center}

\noindent
\normalsize
We study the prospects for HERA with polarized electron and/or proton 
beams to determine the functions describing the fragmentation of a 
longitudinally polarized parton into a polarized $\Lambda$ baryon.
We examine both semi-inclusive deep-inelastic scattering (SIDIS) and 
photoproduction of $\Lambda$.

\vspace{7.0cm}
\noindent
{\sf Contribution to the proceedings of the 1997 workshop on 
`Physics with Polarized Protons at HERA', DESY-Hamburg and
DESY-Zeuthen, March-September 1997.} 
\vfill

\setcounter{page}{0}
\newpage
\begin{center}
{\LARGE \bf Polarized Lambda Production at HERA }

\vspace{1cm}
{\Large D.\ de Florian$^a$, M.\ Stratmann$^b$, W.\ Vogelsang$^a$ }

\vspace*{1cm}
{\it $^a$Theoretical Physics Division, CERN, CH-1211 Geneva 23, Switzerland \\ 
$^b$Department of Physics, University of Durham, Durham DH1 3LE, England}

\vspace*{0.7cm}

\end{center}

\begin{abstract}
We study the prospects for HERA with polarized electron and/or proton 
beams to determine the functions describing the fragmentation of a 
longitudinally polarized parton into a polarized $\Lambda$ baryon.
We examine both semi-inclusive deep-inelastic scattering (SIDIS) and 
photoproduction of $\Lambda$.
\end{abstract}

\section{Introduction}
\noindent
Measurements of rates for single-inclusive $e^+e^-$ annihilation (SIA) into 
a specific hadron $H$,
\begin{equation}
e^+e^- \rightarrow (\gamma,\,Z) \rightarrow H\;X\;\;\;,
\end{equation}
play a similarly fundamental role as those of the corresponding crossed 
`space-like' deep-inelastic scattering (DIS) 
process $ep\rightarrow e' X$. Their 
interpretation in terms of scale-dependent
fragmentation functions $D_f^{H}(z,Q^2)$, the `time-like' counterparts of the 
parton distribution functions $f_H(x,Q^2)$ of a hadron $H$, hence provides
a further important, complementary test of perturbative QCD.
 
The production of $\Lambda$ baryons appears to be particularly  
interesting, also from a different point of view. Contrary to spinless mesons 
like pions and kaons, the $\Lambda$ baryon offers the rather unique 
possibility to study for the first time spin transfer reactions.
The self-analyzing properties of its dominant weak decay 
$\Lambda \rightarrow p \pi^-$ and the particularly large asymmetry of the 
angular distribution of the decay proton in the $\Lambda$ rest-frame 
allow for an experimental reconstruction of the $\Lambda$ spin. 
Such studies of $\Lambda$ polarization could provide a completely new insight 
into the field of spin physics whose theoretical understanding is still far 
from being complete despite of recent progress, and they might also 
yield further information on the hadronization mechanism.

In \cite{burkjaf} a strategy was proposed for extracting in $e^+e^-$
annihilation the functions $\Delta D_f^{\Lambda}(z,Q^2)$ describing 
the fragmentation of a longitudinally polarized parton into a longitudinally 
polarized $\Lambda$~\cite{ji}. At high energies, such as at LEP, 
even {\em{no}} beam polarization is required since the parity-violating 
$q\bar{q}Z$ coupling automatically generates a net polarization of the 
quarks. Here, ALEPH \cite{aleph}, DELPHI \cite{delphi} and OPAL 
\cite{opal} have recently reported first results for 
the polarization of $\Lambda$'s produced on the
`$Z$-resonance'.

On the other hand, after the scheduled upgrade of the HERA electron ring with 
spin rotators in front of the H1 and ZEUS experiments, longitudinally
polarized electrons will be also available for high-energy $ep$ collisions, 
and semi-inclusive DIS (SIDIS) or photoproduction measurements with
polarized $\Lambda$'s in the final state could be carried out.
Furthermore, having also a polarized proton beam available at HERA
would allow the measurement of various different twist-2 asymmetries
depending on whether the $e$ and/or the $p$ beam and/or the
$\Lambda$ are polarized, i.e., $\vec{e}p\rightarrow \vec{\Lambda} X$,
$e \vec{p}\rightarrow \vec{\Lambda} X$, and $\vec{e}\vec{p}\rightarrow 
\Lambda X$ (as usual, an arrow denotes a polarized particle). 

So far estimates for future $\Lambda$ experiments have relied on simple 
models~\cite{nzar} or on Monte-Carlo simulations  
tuned with several parameters and parametrizations of 
scale-{\em{in}}dependent spin-transfer coefficients $C_f^{\Lambda}$ which 
link longitudinally polarized and unpolarized fragmentation functions 
via \cite{bravar}
\begin{equation}
\label{eq:ctrans}
\Delta D_f^{\Lambda}(z) = C_f^{\Lambda}(z) D_f^{\Lambda}(z)\;\;\;.
\end{equation}
Of course, such relations cannot in general hold true in QCD. Due to the 
different $Q^2$-evolutions of $\Delta D_f^{\Lambda}$ and $D_f^{\Lambda}$ the 
assumed scale independence of $C_f^{\Lambda}$ in (\ref{eq:ctrans}) cannot be 
maintained at different values of $Q^2$, and therefore one has to specify a 
scale at which one implements such an ansatz.
 
In the following we will address the issue of fragmentation into polarized
$\Lambda$'s within a detailed QCD analysis. Here it will be possible
for us to work even at next-to-leading order (NLO) accuracy, as the required 
spin-dependent `time-like' 
two-loop evolution kernels were derived recently \cite{tlpol}. For the 
first time, we will provide some realistic sets of unpolarized and polarized 
fragmentation functions for $\Lambda$ baryons. Since there are hardly 
any data sensitive to the polarized fragmentation 
functions $\Delta D_f^{\Lambda}$,
one has to rely on reasonable assumptions. A useful constraint when 
constructing models is provided by the positivity condition (similarly to the 
`space-like' case), i.e.
\begin{equation}
\label{eq:positivity}
\left|\Delta D_f^{\Lambda}(z,Q^2)\right| \le  D_f^{\Lambda}(z,Q^2)\;\;\;.
\end{equation}
The unpolarized $D_f^{\Lambda}(z,Q^2)$ will be taken from an analysis of
the unpolarized case where several different and rather precise measurements 
already exist \cite{newdata}. Of course, it will turn out that the available 
sparse data from LEP \cite{aleph,delphi,opal} are by far not 
sufficient to completely pin down 
the $\Delta D_f^{\Lambda}(z,Q^2)$. Our various proposed sets for 
$\Delta D_f^{\Lambda}$, which are all compatible with the LEP 
data \cite{aleph,delphi,opal}, are 
particularly suited for estimating the physics potential of HERA to 
determine the polarized fragmentation functions 
more precisely. We hence present 
detailed predictions for future SIDIS and photoproduction measurements 
at  HERA using our $\Delta D_f^{\Lambda}$. 

The remainder of this contribution is organized as follows: 
in the next section we briefly sketch the required formalism for 
unpolarized SIA and discuss our analysis of unpolarized
leading order (LO) and NLO $\Lambda$ fragmentation functions. In section 
3 we turn to the case of longitudinally polarized $\Lambda$ production and 
present our different conceivable scenarios for the $\Delta D_f^{\Lambda}$. 
In section 4 we analyze in detail possible SIDIS and photoproduction 
measurements at HERA as a way to further discriminate between the different 
proposed sets of polarized fragmentation functions. Finally our results are 
summarized in section 5. Further details of our analysis will be
published in \cite{paperlambda}.  
%
\section{Unpolarized $\Lambda$ Fragmentation Functions}
\noindent
The cross section for the inclusive production of a hadron with 
energy $E_H$ in SIA at a c.m.s. energy $\sqrt{s}$, integrated over the 
production angle, can be written in the following 
way~\cite{altarelli}:
\begin{equation}
\label{eq:cross}
\frac{1}{\sigma_{tot}} \frac{d\sigma^H}{dx_E} = \frac{1}{\sum_q e_q^2}
\left[ 2\, F_1^{H}(x_E,Q^2) +  F_L^{H}(x_E,Q^2) \right] \; ,
\end{equation}
where $x_E=2 p_H\cdot k/Q^2 = 2 E_H/\sqrt{s}$ ($k$ being the 
momentum of the intermediate boson, $k^2=Q^2=s$) and
$\sigma_{tot}$ is the the total cross section for 
$e^+e^- \rightarrow hadrons$. The sum in (\ref{eq:cross}),
runs over the $n_f$ active quark flavours $q$, and 
the $e_q$ are the corresponding appropriate electroweak charges.

The unpolarized `time-like' structure function $F_1^{H}$ is in LO related 
to the fragmentation functions $D_f^H(z,Q^2)$ by 
\begin{eqnarray}
2 \, F_1^{H}(x_E,Q^2) &=& \sum_q e_q^2 \left[ D_q^H (x_E,Q^2) + 
D_{\bar q}^H (x_E,Q^2) \right]\;\;\; ,  
\end{eqnarray}
the $D_f^H (z,Q^2)$ obeying LO Altarelli-Parisi-type $Q^2$-evolution 
equations~\cite{altarelli}. The corresponding NLO expressions for 
$F_1^H$ and $F_L^H$ which include the relevant NLO coefficient functions 
are too lengthy to be given here but can be found in \cite{altarelli}.
Needless to say that when going to NLO, the $Q^2$-evolution of the
$D_f^H$ also has to be performed to NLO accuracy. The unpolarized two-loop
`time-like' splitting functions required here have been published 
in \cite{split,grvgam}. It should be mentioned that our NLO distributions 
refer to the $\overline{\mathrm{MS}}$ scheme.

In the last few years several experiments \cite{newdata} have reported 
measurements of the unpolarized cross section for the production of 
$\Lambda$ baryons, which allows an extraction of the unpolarized $\Lambda$ 
fragmentation functions required for constructing the polarization 
asymmetries and as reference distributions in the positivity constraint 
(\ref{eq:positivity}). We emphasize at this point that the wide range 
of c.m.s.\ energies covered by the data \cite{newdata} 
($14 \leq \sqrt{s} \leq 91.2$ GeV) makes a detailed QCD analysis that 
includes the $Q^2$-evolution of the fragmentation functions mandatory.
As pointed out in~\cite{burkjaf}, the QCD formalism is strictly speaking
only applicable to strongly produced $\Lambda's$. A certain fraction of the
data~\cite{newdata} will, however, consist of 
secondary $\Lambda$'s resulting 
from $e^+ e^- \rightarrow \Sigma^0 X$ with the subsequent decay 
$\Sigma^0 \rightarrow \Lambda \gamma$, not to be included in the 
fragmentation functions \cite{burkjaf}. For simplicity, we will ignore 
this problem and 
(successfully) attempt to describe the full data samples by fragmentation 
functions that are evolved according to the QCD $Q^2$-evolution equations.

Unless stated otherwise, we will refer to both $\Lambda^0$ and 
$\bar{\Lambda}^0$, which are not usually distinguished in present $e^+e^-$ 
experiments \cite{newdata}, as simply `$\Lambda$'. 
As a result, the obtained fragmentation functions always correspond to the sum
\begin{equation}
\label{eq:lamlambar}
D_f^\Lambda(x_E,Q^2) \equiv D_f^{\Lambda^0}(x_E,Q^2)+ 
D_f^{\bar{\Lambda}^0}(x_E,Q^2)\;\;\;.
\end{equation}
This also considerably simplifies the analysis, since no distinction between 
`favored'  and `unfavored' distributions is required. Since no precise 
SIDIS data are available yet, it is not possible to obtain individual 
distributions for all the light flavours separately, and hence some sensible 
assumptions concerning them have to be made. Employing naive quark model 
$SU_f(3)$ arguments and neglecting any mass differences between 
$u,d,s$, we {\it assume}\footnote{Fits allowed to be more general 
do not significantly improve the final $\chi^2/d.o.f.$} 
that all the light flavours fragment equally into $\Lambda$, i.e.
\begin{eqnarray}
\label{eq:ansatz}
D^{\Lambda}_u=D^{\Lambda}_d=D^{\Lambda}_s=D^{\Lambda}_{\bar{u}}=
D^{\Lambda}_{\bar{d}}=D^{\Lambda}_{\bar{s}} 
\equiv D^{\Lambda}_q\;\;.
\end{eqnarray} 
Needless to say that the $q$ and $\bar{q}$ fragmentation functions 
in (\ref{eq:ansatz}) are equal due to eq.\ (\ref{eq:lamlambar}).  
%
%
\begin{figure}[th]
\begin{center}
\vspace*{-1cm}
\epsfig{file=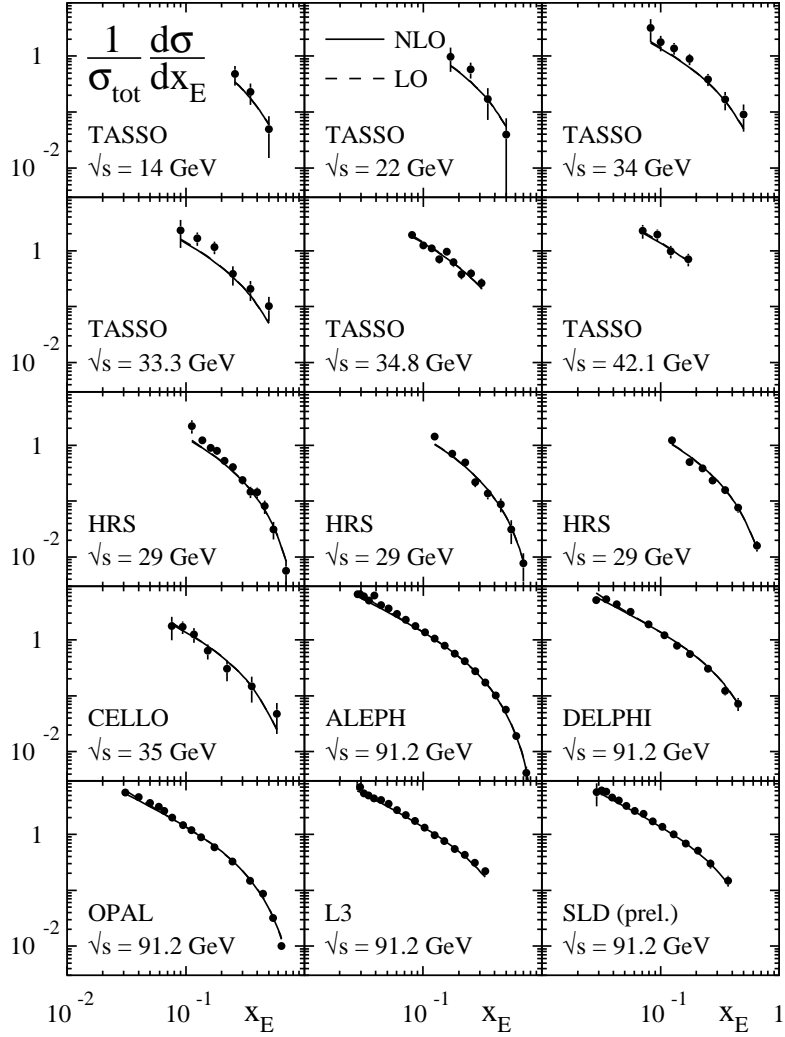,width=12cm}
\vspace*{-1.1cm}
\caption{\sf Comparison of LO and NLO results with all available data on 
unpolarized $\Lambda$ production in $e^+ e^-$ annihilation~\cite{newdata}. 
Note that only data points with $x_E\ge 0.1$ have been included in our fit.}
\vspace*{-0.5cm}
\end{center}
\end{figure}

For our analysis, we choose to work in the framework of the radiative 
parton model
which is characterized by a rather low starting scale $\mu$ for the 
$Q^2$-evolutions. The radiative parton model has proven phenomenologically 
successful in the `space-like' case for both unpolarized~\cite{grv} and 
polarized~\cite{grsv} parton densities, and also in the `time-like' 
situation for photon fragmentation functions~\cite{grvgam,opalg}.

At the initial scale ($\mu^2_{LO}=0.23\,{\mathrm{GeV}}^2$, 
$\mu^2_{NLO}=0.34\,{\mathrm{GeV}}^2$) we choose the following simple ansatz:
\begin{equation}
\label{eq:input}
z\, D^{\Lambda}_f (z,\mu^2) = N_f\, z^{\alpha_f} (1-z)^{\beta_f} \; ,
\end{equation}
where $f=u,d,s$, and $g$. For details on the treatment of the 
heavy flavour contributions, i.e.\ by $c$ and $b$ quarks, which are
of minor importance in our analysis, we refer to \cite{paperlambda}.
Utilizing eq.\ (\ref{eq:ansatz}) a total of 10 free parameters remains 
to be fixed from a fit to the available 103 data 
points \cite{newdata} (as usual, see, e.g. \cite{bkk} for a recent
analysis of pion and kaon data, we include only data with $x_E>0.1$
in our fit to avoid possibly large non-perturbative contributions
induced by finite-mass corrections). The total 
$\chi^2$ values are 103.55 and 104.29 in NLO and LO, respectively, and 
the optimal parameters in (\ref{eq:input}) can be 
found in~\cite{paperlambda}. 

A comparison of our LO and NLO results with the data is presented in fig.\ 1, 
where all the existing data \cite{newdata}
have been converted to the `format' of 
eq.(\ref{eq:cross}). One should note that the LO and the NLO results are 
almost indistinguishable, demonstrating the perturbative stability of
the process considered. Furthermore, there is an excellent agreement between 
the predictions of our fits and the data even in the small-$x_E$ region 
which has not been included in our analysis.
%
\section{Polarized Fragmentation Functions}
%
\noindent 
Having obtained a reliable set of unpolarized fragmentation 
functions we now turn to the polarized case where unfortunately
only scarce and far less precise data are available. In fact, no data at 
all have been obtained so far using polarized beams. The
only available information comes from {\it un}polarized LEP measurements 
\cite{aleph,delphi,opal} profitting from the parity-violating 
electroweak $q\bar{q}Z$ coupling. 

For such measurements, done at the mass of the $Z$ boson (`$Z$-resonance'), 
the cross section for the production of polarized hadrons can be written 
as \cite{burkjaf,ravi}
\begin{equation}
\label{eq:pol}
\frac{d\Delta \sigma^H}{d\Omega dx_E} = 3 \frac{\alpha^2(Q^2)}{2 s}\left[  
g_3^H(x_E,Q^2) (1+\cos^2\theta) -  g_1^H(x_E,Q^2) \cos\theta  
+ g_L^H(x_E, Q^2) (1- \cos^2\theta) \right] \; .
\end{equation}
If, as for the quoted experimental results~\cite{aleph,delphi,opal}, the 
cross section is integrated over the angle $\theta$, the anyway small 
contribution from $g_1^H$ vanishes. One can then define the asymmetry
\begin{equation}
\label{eq:lepasym}
A^H = \frac{ g_3^H + g_L^H/2}{ F_1^H+F_L^H/2}
\end{equation}
which corresponds to the `$\Lambda$-polarization' observable measured at LEP.
The polarized {\it non}-singlet structure function $g_3^H$ is given at LO by 
\begin{equation}
g_3^H(x_E,Q^2) = \sum_q \, g_q \left[
\Delta D_q^H (x_E,Q^2) -\Delta D_{\bar q}^H (x_E,Q^2) \right]\;\;\;,    
\end{equation}
where the $g_q$ are the appropriate effective charges 
(see, e.g., \cite{paperlambda}). The complete NLO QCD corrections 
can be found in \cite{paperlambda,ravi}. 
Note that only the {\it valence} part 
of the polarized fragmentation functions can be obtained from the available 
LEP data~\cite{aleph,delphi,opal}, and that $\Lambda^0$'s and 
$\bar{\Lambda}^0$'s give contributions of opposite signs to the measured 
polarization and thus to $g_3^\Lambda$. Unfortunately, 
it turns out that with the available LEP data~\cite{aleph,delphi,opal}, 
all obtained on the $Z$-resonance, it is not 
even possible to obtain the valence distributions for all the flavours, 
so some assumptions have to be made here. Obviously, even further 
assumptions are needed for the polarized gluon and sea fragmentation 
functions in order to have a complete set of fragmentation functions 
suitable for predictions for other processes, in particular for those relevant
for HERA.

The heavy flavour contributions to polarized $\Lambda$ 
production are neglected, and $u$ and $d$ fragmentation functions 
are taken to be equal in this analysis. Furthermore, polarized  
unfavored distributions, i.e. $\Delta D_{\bar{u}}^{\Lambda^0} = 
\Delta D_{u}^{\bar{\Lambda}^0}$, etc., and the gluon fragmentation 
function $\Delta D_g^\Lambda$ are assumed to be negligible at the 
initial scale $\mu$, an assumption which of course might deserve a further
scrutiny in the future. The remaining fragmentation functions are 
parametrized in the following simple way
\begin{equation}
\label{eq:polinput}
\Delta D_s^\Lambda (z,\mu^2) = z^\alpha  D_s^\Lambda (z,\mu^2)\;\;,\;\;
\Delta D_u^\Lambda (z,\mu^2) = \Delta  D_d^\Lambda (z,\mu^2) = 
N_u\, \Delta D_s^\Lambda (z,\mu^2) 
\end{equation}
and are subject to the positivity constraints (\ref{eq:positivity}).
These input distributions are then evolved to higher $Q^2$ via the 
appropriate Altarelli-Parisi equations. For the NLO evolution one
has to use for this purpose the spin-dependent `time-like' two-loop 
splitting functions as derived in~\cite{tlpol} in the $\overline{\mathrm{MS}}$
scheme. 
%
%
\begin{figure}[th]
\begin{center}
\vspace*{-1.4cm}
\epsfig{file=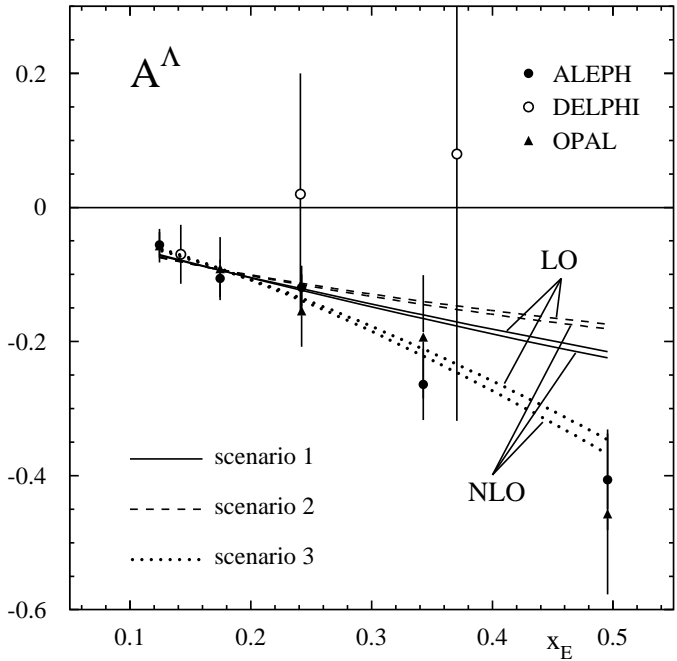,width=10.5cm}
\vspace*{-1.1cm}
\caption{\sf Comparison of LEP data \cite{aleph,delphi,opal} and our LO 
and NLO results for the asymmetry $A^{\Lambda}$, using the three different
scenarios (see text).}
\vspace*{-0.65cm}
\end{center}
\end{figure} 

Within this framework we try three different scenarios 
for the polarized fragmentation functions at our low
initial scale $\mu$ to cover a rather wide range of plausible models: 

\noindent
{\bf{Scenario 1}} corresponds to the expectations from the non-relativistic 
naive quark model where only $s$-quarks can contribute to the fragmentation
processes that eventually yield a polarized $\Lambda$, even if the $\Lambda$
is formed via the decay of a heavier hyperon. We hence have $N_u=0$ in 
(\ref{eq:polinput}) for this case.

\noindent
{\bf{Scenario 2}} is based on estimates by Burkardt and 
Jaffe \cite{burkjaf,jaffe2} for the DIS structure function 
$g_1^{\Lambda}$ of the $\Lambda$, predicting
sizeable negative contributions from $u$ and $d$ quarks to $g_1^{\Lambda}$ by 
analogy with the breaking of the Gourdin-Ellis-Jaffe sum rule \cite{gej}
for the proton's $g_1^p$. Assuming that such features also carry over
to the `time-like' case \cite{jaffe2}, we simply impose $N_u=-0.20$ 
(see also \cite{bravar}).

\noindent
{\bf{Scenario 3:}} All the polarized fragmentation functions are assumed 
to be equal here, i.e. $N_u=1$, contrary to the expectation  
of the non-relativistic quark model used in scen.\ 1. This rather `extreme' 
scenario might be realistic if, e.g., there are sizeable contributions to
polarized $\Lambda$ production from decays of heavier hyperons who have
inherited the polarization of originally produced $u$ and $d$ quarks.

Our results for the asymmetry $A^{\Lambda}$ in (\ref{eq:lepasym}) within 
the three different scenarios are compared to the available LEP 
data~\cite{aleph,delphi,opal} in fig.\ 2.
The optimal parameters in (\ref{eq:polinput}) for the three models can
be found in \cite{paperlambda}. As can be seen, the best agreement with 
the data is obtained within the (naively) most unlikely scen.\ 3.  
The differences occur mainly in the region of large $x_E$, where 
scen.\ 1 and 2 cannot fully account for the rather large 
observed polarization 
due to the positivity constraints (\ref{eq:positivity}). 
For instance, in the case of scen.\ 1, the 
asymmetry behaves asymptotically roughly like $-\Delta D_s^\Lambda / 3D_s^\Lambda$, 
and even when saturating the positivity constraint (\ref{eq:positivity}) 
at around $x_E=0.5$ it is not possible to obtain a polarization as 
large as the one required by the LEP data. Of course, such an 
argument still depends strongly on the assumed $SU(3)_f$ symmetry for 
the {\it{un}}polarized fragmentation functions, which could be broken. The 
situation concerning the $\Lambda$ fragmentation functions can only be 
further clarified by future precise SIDIS and photoproduction measurements 
for unpolarized {\em and} polarized beams.
%
\begin{figure}[th]
\begin{center}
\vspace*{-1cm}
\epsfig{file=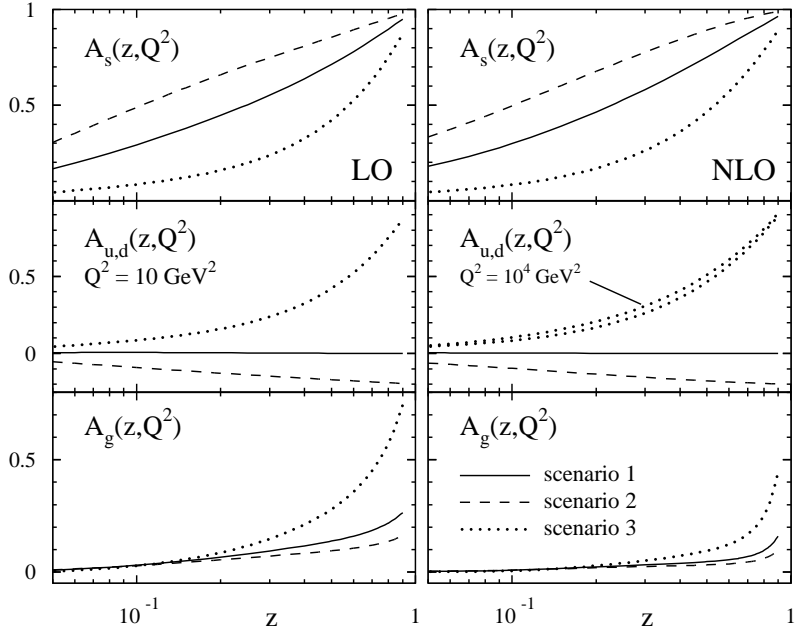,width=12cm}
\vspace*{-1.1cm}
\caption{\sf LO and NLO partonic fragmentation asymmetries $A_f\equiv\Delta 
D^{\Lambda}_f/D^{\Lambda}_f$ for $f=s$, $u=d$, and $g$ at 
$Q^2=10\,{\mathrm{GeV}}^2$. For the NLO $A_{u,d}$ we also show the results 
for $Q^2=10^4\,{\mathrm{GeV}}^2$.}
\vspace*{-0.6cm}
\end{center}
\end{figure}

Finally, in fig.\ 3 we show the LO and NLO partonic fragmentation 
asymmetries for 
each flavour distribution separately, i.e., $A_f\equiv\Delta D^{\Lambda}_f/
D^{\Lambda}_f$. A rather large polarized gluon fragmentation function
has built up in the $Q^2$-evolution in spite of the
vanishing input at $\mu^2$. 
\section{SIDIS and Photoproduction Signatures at HERA}
\noindent
Equipped with various sets of polarized fragmentation 
functions, let us now turn to other processes that might further probe our 
distributions.

The SIDIS process $e N \rightarrow e' H X$ is also very well suited to 
give information on fragmentation functions. In this case, the cross 
section is proportional to a combination of both the parton distributions 
of the nucleon $N$ and the fragmentation functions for the hadron $H$.
The latter thus automatically appear in a constellation different from the 
one probed in $e^+ e^-$ annihilation. 
 
In this work, we will only refer to the current fragmentation region by 
implementing a cut $x_F>0$ on the Feynman-variable. Target fragmentation can 
be accounted for by the introduction of 
`fracture functions' \cite{veneziano}, 
but is beyond the scope of this analysis \cite{ellis}.
%
%
\begin{figure}[th]
\begin{center}
\vspace*{-2cm}
\hspace*{0.8cm}
\epsfig{file=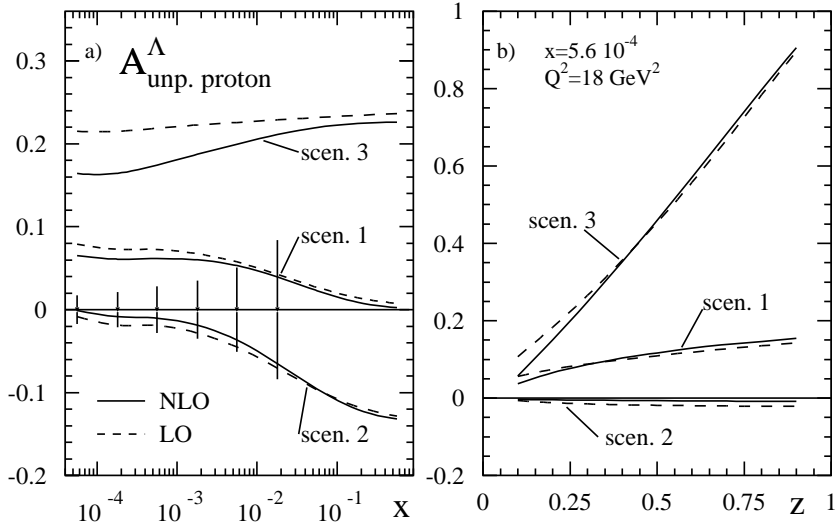,width=10.5cm}
\vspace*{-1.6cm}
\caption{\sf LO and NLO predictions for the SIDIS asymmetry for 
unpolarized protons and polarized $\Lambda$'s and leptons 
(see text) for our three
distinct scenarios of polarized fragmentation functions. In {\bf a)} we also
show the expected statistical errors for such a measurement at HERA.}
\vspace*{-0.6cm}
\end{center}
\end{figure}

In the particular case where both nucleon and hadron are unpolarized, the 
cross section can be written in a way similar to the fully inclusive DIS 
case:
\begin{equation}
\label{eq:sidis}
\frac{d\sigma^H}{dx\, dy\, dz} = 
\frac{4\, \pi\alpha^2 x s}{Q^4} 
\left[ (1+(1-y)^2) F_1^{N/H}(x,z,Q^2) + 
\frac{(1-y)}{x} F_L^{N/H}(x,z,Q^2) \right] 
\end{equation}
with the structure function $F_1^{N/H}$ given at LO by
\begin{equation}
\label{eq:nopol} 
2\, F_{1}^{N/H}(x,z,Q^2) = \sum_{q,\overline{q}} 
e_q^2 \, q (x,Q^2)  D^H_q (z,Q^2)  \; . 
\end{equation}
The corresponding NLO corrections can be found in \cite{graudenz,altarelli}.
As already mentioned in the introduction, three other possible cross
sections can be defined when the polarization of the lepton, the initial 
nucleon and the hadron are taken into account. 
If both nucleon and hadron are polarized and the lepton is unpolarized, 
the expression is similar to eqs.\ (\ref{eq:sidis}),(\ref{eq:nopol}) 
above with, however, the unpolarized parton distributions {\em and} the 
fragmentation functions to be replaced by their polarized counterparts. 
In the case that the lepton and either 
the nucleon or the hadron are polarized, the expression for the cross 
section is given as in the fully inclusive case by a single structure 
function $g_{1}^{N/H}(x,z,Q^2)$: 
\begin{equation}
\label{eq:sidispol}
\frac{d\Delta\sigma^H}{dx\, dy\, dz} = \frac{8 \pi\alpha^2 x y s}{Q^4} 
\left[ (2-y)\, g_1^{N/H}(x,z,Q^2)  \right]  \; .
\end{equation}
Here the polarized structure function $g_1^{N/H}$ can be written as
\cite{singlepol,doublepol}
\begin{equation}
\label{eq:pol1} 
2\, g_{1}^{N/H}(x,z,Q^2) = \sum_{q,\overline{q}} 
e_q^2 \, (\Delta) q (x,Q^2) (\Delta) D^H_q (z,Q^2)  \; , 
\end{equation}
the position of the $\Delta$ depending on which particle is polarized.
The corresponding NLO corrections for the various polarized cross
sections can be found in \cite{singlepol,doublepol,thesis,paperlambda}.
%
%
\begin{figure}[th]
\begin{center}
\vspace*{-2cm}
\epsfig{file=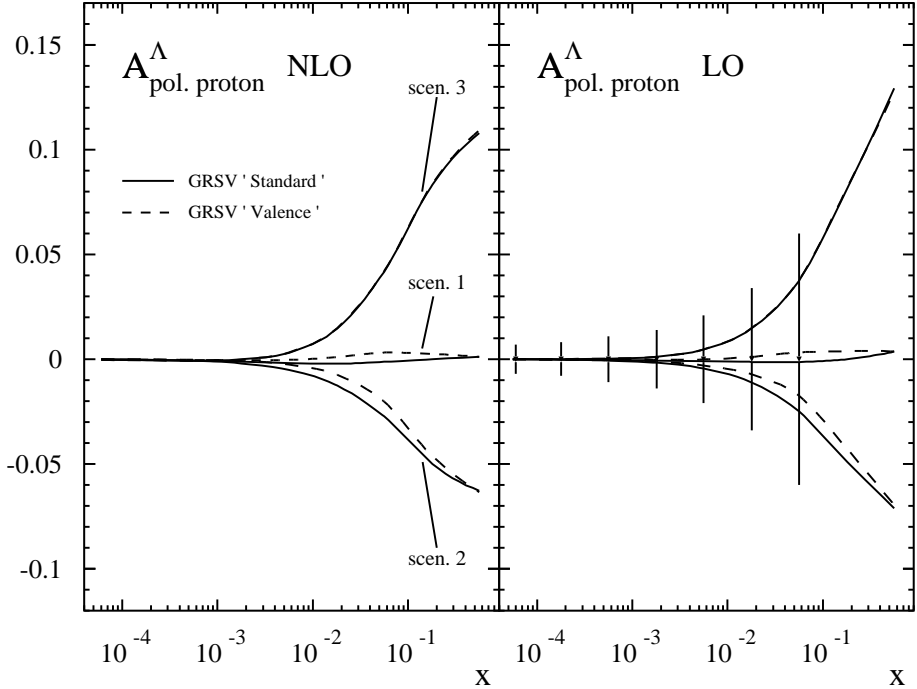,width=12.5cm}
\vspace*{-1.6cm}
\caption{\sf LO and NLO predictions for the SIDIS asymmetry for polarized
protons but unpolarized leptons for two different sets of polarized
parton distributions taken from \cite{grsv}. Also shown are the
expected statistical errors for such a measurement at HERA assuming
again a luminosity of $100\,{\mathrm{pb}}^{-1}$ and a $\Lambda$
detection efficiency of 0.1.}
\vspace*{-0.6cm}
\end{center}
\end{figure}

The most interesting observable at HERA with respect to the determination
of the polarized $\Lambda$ fragmentation functions is of course the 
asymmetry for the production of {\em polarized} $\Lambda$'s from an 
unpolarized proton, defined by $A^{\Lambda}\equiv g_1^{p/\Lambda}/
F_1^{p/\Lambda}$ \cite{jaffe2}. Such a measurement 
would be particularly suited 
to improve the information on polarized fragmentation functions. 
In fig.\ 4a), we show our LO and NLO predictions for HERA with polarized 
electrons and {\it{un}}polarized protons using the GRV parton 
distributions \cite{grv}, integrated over the measurable range 
$0.1 \leq z \leq 1$. The values for $Q^2$ that correspond to each $x$ have 
been chosen as in \cite{desh}. Good perturbative stability of 
the process is found. As can be seen, the results obtained using the 
three distinct scenarios for polarized fragmentation functions turn 
out to be completely different. Since the asymmetry at small 
$x$ is determined by the proton's sea quarks, its behaviour can 
be easily understood: 
in scen.\ 1 only $s$ quarks fragment to polarized $\Lambda$'s, 
giving an asymmetry which is positive but about three times 
smaller than the one of scen.\ 3 where all the flavours contribute. 
In the case of scen.\ 2 the positive contribution from $s$-quark 
fragmentation is cancelled by a negative one from $u$ and $d$, resulting 
in an almost vanishing asymmetry. The interpretation is similar for 
the region of large $x$, where only the contribution involving $u_{v}$ 
is sizeable and the asymmetry 
asymptotically goes to $\int \,dz\Delta D_u^\Lambda / \int \,dz 
D_u^\Lambda$ for each scenario.

We have included in fig.~4a) also the expected statistical errors 
for HERA, computed assuming 
an integrated luminosity of $500$ pb$^{-1}$ and a realistic value of
$\epsilon=0.1$ for the efficiency of $\Lambda$ detection \cite{albert}. 
Comparing the asymmetries and the error bars in fig.~4a) one concludes that a 
measurement of $A^{\Lambda}$ at small $x$ would allow a discrimination 
between different conceivable scenarios for polarized fragmentation functions.
Fig.~4b) shows our results vs.\ $z$ for fixed $x=5.6 \cdot 10^{-4}$. Again, 
very different asymmetries are found for the three scenarios. 

The particular case of both target and hadron being polarized was originally 
proposed as a very good way to obtain the $\Delta s$ distribution~\cite{luma}. 
The underlying assumption here was that only the fragmentation function 
$\Delta D_s^{\Lambda}$ is sizeable (as realized, e.g. in our scenario 1), and 
that therefore the only contribution to the polarized cross section has to be 
proportional to $\Delta s \Delta D_s^{\Lambda}$. In order to analyse the 
sensitivity of the corresponding asymmetry to $\Delta s$, we compute it 
using the two different GRSV sets of polarized parton densities of the 
proton \cite{grsv}, which mainly differ in the strange distribution:
the so-called `standard' set assumes an unbroken $SU(3)_f$ symmetric
sea, whereas in the `valence' scenario the sea is maximally broken
and the resulting strange quark density quite small.

The results are shown in fig.\ 5. Unfortunately -- and not unexpectedly --
it turns out that the differences in the asymmetry resulting from our 
different models for polarized $\Lambda$ fragmentation are far larger 
than the ones due to employing different polarized proton strange densities.
In addition, a distinction between different $\Delta s$ would remain 
elusive even if the spin-dependent $\Lambda$ fragmentation functions
were known to good accuracy, as can be seen from the error bars
in fig.~5 which were obtained using the same parameters as before.
%
%
\begin{figure}[th]
\begin{center}
\vspace*{-1.2cm}
\hspace*{0.3cm}
\epsfig{file=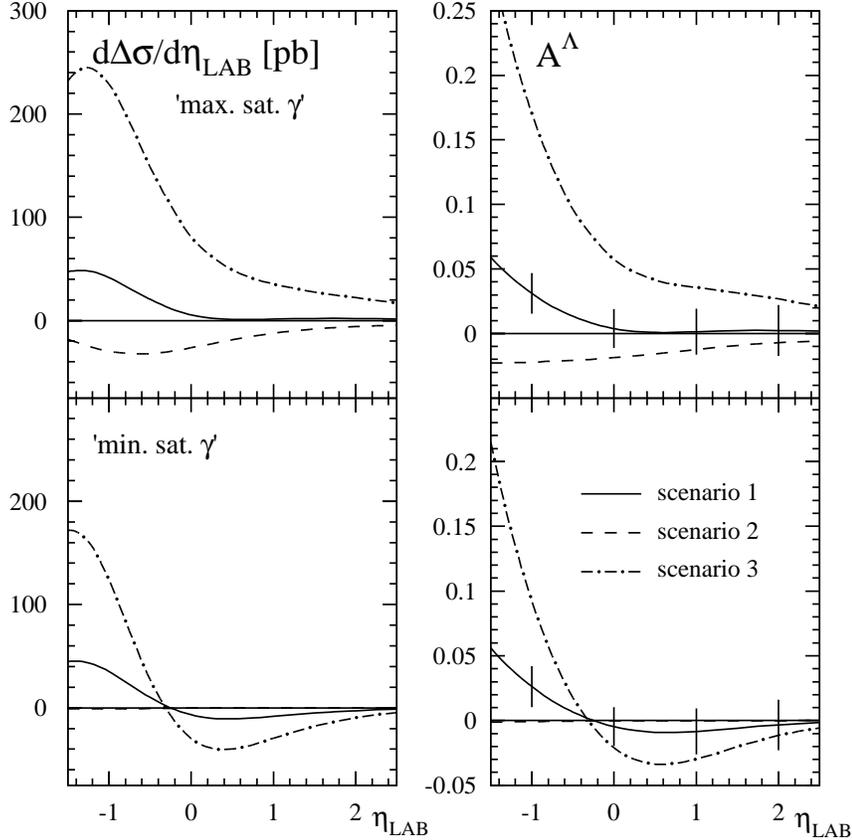,width=12.5cm}
\vspace*{-1.cm}
\caption{\sf $\eta_{LAB}$-dependence of the polarized
single-inclusive $\Lambda$ photoproduction cross section and of
the corresponding asymmetry $A^{\Lambda}$ at HERA, integrated over 
$p_T^{\Lambda}>2\,{\mathrm{GeV}}$. The renormalization/factorization scale was
chosen to be $p_T^{\Lambda}$. In the upper part the resolved contribution to
the cross section has been calculated with the `maximally'
saturated set of polarized photonic parton distributions whereas
the `minimally' saturated one was used in the lower half \cite{photo}. Also
shown are the expected statistical errors for such a measurement
at HERA assuming a luminosity of $100\,{\mathrm{pb}}^{-1}$ and a
$\Lambda$ detection efficiency of 0.1.}
\vspace*{-0.6cm}
\end{center}
\end{figure}

Finally, let us examine the case of photoproduction of $\Lambda$'s
at HERA. We can be very brief here and refer the reader to the dedicated
report on photoproduction for this workshop \cite{photo}, where all
details like, e.g. the polarized parton distributions of the photon
and the kinematical cuts suitable for photoproduction, are discussed.

Here it appears very promising to study the $\eta_{LAB}$-distribution
of the photoproduction cross section for $\Lambda$'s 
and the corresponding asymmetry, where
$\eta_{LAB}$ is the laboratory frame rapidity defined to be positive
in the proton forward direction. Fig.\ 6 shows our results for such a
calculation where we have integrated over $p_T^{\Lambda}>2\,{\mathrm{GeV}}$.
To calculate the resolved photon contribution to the cross section
we have used two conceivable, extreme scenarios for the 
experimentally completely unknown parton distributions of a longitudinally
polarized photon $\Delta f^{\gamma}$ (see \cite{photo} for details).
As can be inferred from fig.\ 6 there is a strong sensitivity to
our proposed different scenarios for the $\Delta D_f^{\Lambda}$, but
a comparison of the upper and lower parts of fig.\ 6, which have been 
calculated for different assumptions for the resolved contribution, also 
reveals that our ignorance concerning the $\Delta f^{\gamma}$ sets a 
potential limitation to an extraction of the fragmentation functions.
The $\Delta f^{\gamma}$ can of course be further constrained in 
other conceivable photoproduction measurements in polarized $ep$ collisions
at HERA like dijet production \cite{photo}. Finally it should be 
stressed that due to the small $Q^2$, photoproduction benefits from 
larger rates, and hence reasonable measurements can be performed at even 
rather low luminosities of about $100\,{\mathrm{pb}}^{-1}$.
%
\section{Summary and Conclusions}
\noindent
We have analyzed HERA's capability of pinning down the spin-dependent 
fragmentation functions for $\Lambda$ baryons via 
semi-inclusive deep-inelastic scattering or photoproduction of $\Lambda$'s. 

Working within the framework of the radiative parton model, our starting 
point has been a fit to unpolarized data for $\Lambda$ production taken
in $e^+ e^-$ annihilation, yielding a set of realistic unpolarized 
fragmentation functions for the $\Lambda$. We have then made simple 
assumptions for the relation between the spin-dependent and the unpolarized 
$\Lambda$ fragmentation functions at the input scale for the $Q^2$ evolution. 
Taking into account the sparse LEP data on $\Lambda$ polarization,
we were able to set up three distinct `toy scenarios' for the spin-dependent
$\Lambda$ fragmentation functions. We emphasize that our proposed sets can by
no means cover all the allowed possibilities for the polarized fragmentation 
functions, the main reason being that the LEP data are only sensitive
to the valence part of the polarized fragmentation functions. Thus, there are 
still big uncertainties related to the `unfavored' quark and gluon 
fragmentation functions, making further measurements in other processes
indispensable.

Under these premises, we have considered $\Lambda$ production at HERA, both
in a situation with polarized electrons scattered of unpolarized protons
and vice versa. It turns out that SIDIS measurements in the first case
should be particularly well suited to yield further information on the $\Delta 
D_f^{\Lambda}$: differences between the asymmetries obtained when using
different sets of $\Delta D_f^{\Lambda}$ are usually larger than the
expected statistical errors. In contrast to this, having a polarized proton 
beam does not appear beneficial in any way as far as $\Lambda$ production
at HERA is concerned. Finally, we found that also photoproduction of 
$\Lambda$'s in scattering of polarized photons off the unpolarized proton
beam at HERA shows strong sensitivity to the $\Delta D_f^{\Lambda}$, even 
though here the unknown contribution from resolved photons will set
limitations to an extraction of the polarized fragmentation functions.  

\section*{Acknowledgements}
The work of one of us (DdF) was partially supported by the World Laboratory.
%

%
\end{document}